\documentclass
[superscriptaddress,secnumarabic,amssymb,amsmath,nobibnotes,aps,prd,showkeys,nofootinbib,nopacsnumber,onecolumn,10pt]{revtex4}%
\usepackage{graphicx}
\usepackage{bm}
\usepackage{amsmath}
\usepackage{amsfonts}
\usepackage{amssymb}
\usepackage{tikz}%
\providecommand{\U}[1]{\protect\rule{.1in}{.1in}}
\usetikzlibrary{arrows.meta, positioning, shapes.geometric}

\newcommand{\be}{\begin{equation}}
\newcommand{\ee}{\end{equation}}

\newcommand{\mincir}{\raise
-3.truept\hbox{\rlap{\hbox{$\sim$}}\raise4.truept\hbox{$<$}\ }}
\newcommand{\magcir}{\raise
-3.truept\hbox{\rlap{\hbox{$\sim$}}\raise4.truept\hbox{$>$}\ }}

\begin{document}
\title{Canonical Structure and Hidden Symmetries in Scalar Field Cosmology}
\author{Andronikos Paliathanasis}
\email{anpaliat@phys.uoa.gr}
\affiliation{Institute for
Systems Science, Durban University of Technology, Durban 4000, South Africa}
\affiliation{Departamento de Matem\`{a}ticas, Universidad Cat\`{o}lica del Norte, Avda.
Angamos 0610, Casilla 1280 Antofagasta, Chile}
\affiliation{National Institute for Theoretical and Computational Sciences (NITheCS), South Africa.}

\begin{abstract}
We investigate hidden symmetries in a minimally coupled scalar field cosmology
within the FLRW universe, considering a perfect fluid both with and without
interaction with the scalar field. We show that, for an exponential potential,
there exists a set of canonical transformations through which the cosmological
field equations can be recast as those of a free particle in flat space. Based
on this equivalence, we construct a mapping that generates cosmological
solutions with interaction terms, corresponding to a chameleon mechanism.
Finally, we discuss how this class of canonical transformations can relate the
solution spaces of different cosmological models, such as those of the scalar
field and of the $\Lambda$-cosmology.

\end{abstract}
\date{\today}
\maketitle


\section{Introduction}

Exact and analytic solutions are important in all areas of physics. In
gravitational physics and cosmology, such solutions allow for the precise
modeling of gravitational fields and spacetime structures~\cite{ex1,ex2}. In
cosmological studies, they serve as reference points for describing asymptotic
behaviors near singularities and provide examples of the dynamical evolution
of dark energy. For a thorough and still relevant discussion on the subject,
we refer the reader to~\cite{rp1}.

The analysis of the recent cosmological data indicates deviations from
$\Lambda$-Cosmology, suggesting that dark energy may possess a dynamical
structure~\cite{desi1,desi2,desi3,desi4,desi5,desi6,desi7,desi8,desi9,desi10}.
Scalar fields have been employed to describe the early accelerated expansion
of the universe, the inflation
\cite{guth,inf2,ref100,ref4,ref5a,ref7,ref10,orl2,olr3,olr4,pp1}. In addition,
quintessence and phantom fields have been used as simple mechanisms to model
dynamical dark energy \cite{ratra,peebles,dn0}. For quintessence, the
equation-of-state parameter is bounded; however, this is not the case for
phantom fields \cite{q14}, where Big Rip singularities can arise \cite{q15}.
However, these Big Rip singularities can be avoided when a non-zero
interaction is introduced between the scalar field and the source of matter
\cite{q16,q17}. Cosmological interactions have also been introduced to address
the coincidence problem and various cosmological
tensions~\cite{ame00,con1,ht4,ht5,ss4,ss5,ss12,ss16,ss17,ss18,w7,oro1}.

The presence of a nonzero interaction term implies that the mass of the scalar
field is influenced by the mass of the ideal gas, leading to the so-called
chameleon mechanism. In particular, in~\cite{ch1,ch2}, a coupling function was
introduced between the scalar field and the energy density of the fluid,
allowing for energy transfer between the two components~\cite{ame00}. The
coupling term modifies the Klein--Gordon equation for the scalar field by
introducing an effective potential, which leads to an effective mass. This
mechanism allows the scalar field to acquire a large mass in dense
environments, such as near the surface of the Earth, and a small mass in
regions of low energy density, such as in outer space.

A geometric approach to derive such coupling functions is provided by Weyl
Integrable Spacetime \cite{va5}. In this gravitational theory, two conformally
related metrics are defined, with the scalar field acting as the conformal
factor. Although this framework naturally leads to an exponential coupling
function, it can be generalized to other forms by introducing a more general
kinetic term in the gravitational action for the scalar field
\cite{va6,va7,va4,w1,w2}.

Within the Friedmann--Lema\^{\i}tre--Robertson--Walker (FLRW) geometry, the
cosmological field equations for scalar field models remain second-order.
However, due to the nonlinear nature of the equations, only a few exact and
analytic solutions are known in the literature; see, for instance,
\cite{exx1,exx2,exx3,ex4,ex5,ex6,ex7,ex8,Barrow,Urena,l1} and references
therein. Even fewer solutions exist when the chameleon mechanism is included.
Only recently has a family of analytic solutions in chameleon cosmology been
derived using the method of variational symmetries \cite{nsch1}.

In this work, we study the cosmological field equations for scalar field
models with an interaction term using the Eisenhart--Duval lift
\cite{ll01,ll02}. Within the minisuperspace description, we construct a new,
equivalent dynamical system of geodesic equations that shares the same
solution space as the original cosmological field
equations~\cite{ll03,ll05,ll06}. We determine the scalar field potential and
the interaction term such that the field equations become globally
linearizable and equivalent to the equations of motion of a free particle,
thereby revealing hidden symmetries. This property enables us to explore the
solution spaces of different gravitational models and to determine mappings
that transform solutions of one model into those of another. This study not
only provides new directions for investigating the cosmological field
equations but also offers important insights into the nature of the
minisuperspace where the dynamical variables of the cosmological model are
defined. The structure of the paper is as follows.

In Section~\ref{sec2}, we present the basic elements and definitions of scalar
field cosmology within a spatially flat FLRW geometry. The chameleon
mechanism, which describes energy transfer between the scalar field and the
matter source, is introduced in Section~\ref{sec3}. Section~\ref{sec4}
contains the main results of this study, where we investigate the existence of
canonical transformations that relate scalar field models with and without
interaction terms. We apply the Eisenhart-Duval lift approach and define
equivalent extended Hamiltonian systems that share the same solutions as the
cosmological field equations. We show that, for an exponential potential and
either an exponential interaction or no interaction at all, canonical
transformations exist such that the cosmological field equations can be
expressed in the form of a three-dimensional free particle. Consequently, the
solution spaces of scalar field models with and without interaction terms are
equivalent. Finally, in Section~\ref{sec5}, we summarize our conclusions and
visualize this correspondence in a diagram form.

\section{Scalar field Cosmology}

\label{sec2}

In modern cosmology, scalar fields play an important role because they provide
a simple way to introduce new dynamical degrees of freedom into the
gravitational field equations, which are necessary to explain observational
phenomena~\cite{ameb10,soti} and to address the dark energy problem.

We work within the framework of General Relativity. The gravitational field
equations are derived from the gravitational action integral,
\begin{equation}
S=S_{EH}+S_{\phi}+S_{m}, \label{ac.01}%
\end{equation}
where
\begin{equation}
S_{EH}=\int d^{4}x\sqrt{-g}\frac{R}{2}, \label{ac.02}%
\end{equation}
is the Einstein-Hilbert Action, $S_{\phi}$ remarks for the Action Integral for
the canonical scalar field
\begin{equation}
S_{\phi}=-\int\sqrt{-g}d^{4}x\left(  \frac{\varepsilon}{2}g^{\mu\kappa}%
\nabla_{\mu}\phi\left(  x^{\nu}\right)  \nabla_{\kappa}\phi\left(  x^{\nu
}\right)  +V\left(  \phi\left(  x^{\nu}\right)  \right)  \right)  ,
\label{ac.03}%
\end{equation}
and $S_{m}$ describes the matter component, that is,
\begin{equation}
S_{m}=-\int\sqrt{-g}d^{4}x~L_{m}\left(  \psi\left(  x^{\nu}\right)
,\nabla_{\mu}\psi\left(  x^{\nu}\right)  \right)  . \label{ac.04}%
\end{equation}

$R$ is the Ricci scalar of the metric tensor $g_{\mu\nu}$.

We assume that the universe is isotropic and homogeneous, described by the
spatially flat FLRW metric with line element
\begin{equation}
ds^{2}=-N^{2}\left(  t\right)  dt^{2}+a^{2}\left(  t\right)  \left(
dx^{2}+dy^{2}+dz^{2}\right)  , \label{ac.05}%
\end{equation}
where $N\left(  t\right)  $ is the lapse function, and the scale factor
$a\left(  t\right)  $ describes the radius of the three-dimensional
hypersurface. The volume~$V\left(  t\right)  $ of the hypersurface is defined
as $V\left(  t\right)  =a^{3}\left(  t\right)  $.

The parameter $\varepsilon$ in $S_{\phi}$ is constrained such that
$\varepsilon^{2}\rightarrow1$, where for $\varepsilon=+1$, the scalar field
corresponds to quintessence with an equation-of-state parameter
\begin{equation}
w_{\phi}=\frac{g^{\mu\kappa}\nabla_{\mu}\phi\left(  x^{\nu}\right)
\nabla_{\kappa}\phi\left(  x^{\nu}\right)  -2V\left(  \phi\left(  x^{\nu
}\right)  \right)  }{g^{\mu\kappa}\nabla_{\mu}\phi\left(  x^{\nu}\right)
\nabla_{\kappa}\phi\left(  x^{\nu}\right)  +2V\left(  \phi\left(  x^{\nu
}\right)  \right)  }, \label{ac.06a}%
\end{equation}
which is bounded by $\left\vert w_{\phi}\right\vert \leq1$. On the other hand,
the value $\varepsilon=-1$ corresponds to a phantom field, where $w_{\phi}$
can cross the phantom divide line, i.e., it can take values smaller than $-1$.
The equation of state parameter for the phantom field is expressed as
\begin{equation}
w_{\phi}=\frac{-g^{\mu\kappa}\nabla_{\mu}\phi\left(  x^{\nu}\right)
\nabla_{\kappa}\phi\left(  x^{\nu}\right)  -2V\left(  \phi\left(  x^{\nu
}\right)  \right)  }{-g^{\mu\kappa}\nabla_{\mu}\phi\left(  x^{\nu}\right)
\nabla_{\kappa}\phi\left(  x^{\nu}\right)  +2V\left(  \phi\left(  x^{\nu
}\right)  \right)  }. \label{ac.06}%
\end{equation}

The potential function $V\left(  \phi\left(  x^{\nu} \right)  \right)  $
defines the mass of the scalar field and plays an important role in the
physics of the model. For the Lagrangian of the matter component, we assume
that it describes a perfect fluid with energy density $\rho_{m}$, pressure
$p_{m}$, and constant equation-of-state parameter $p_{m} = w_{m} \rho_{m}$,
that is, $L_{m}\left(  \psi\left(  x^{\nu} \right)  , \nabla_{\mu} \psi\left(
x^{\nu} \right)  \right)  = -\rho_{m}$.

Variation of the Action Integral with the metric tensor leads to the
Einstein's field equations%
\begin{equation}
G_{\mu\nu}=T_{\mu\nu}^{\phi}+T_{\mu\nu}^{m}, \label{ac.07}%
\end{equation}
where $T_{\mu\nu}^{\phi}$ and $T_{\mu\nu}^{m}$ are the energy momentum tensors
for the scalar field and the matter components respectively, that is,%
\begin{equation}
T_{\mu\nu}^{\phi}=\varepsilon\nabla_{\mu}\phi\nabla_{\nu}\phi-g_{\mu\nu
}\left(  \frac{\varepsilon}{2}g^{\kappa\lambda}\nabla_{\kappa}\phi
\nabla_{\lambda}\phi+V\left(  \phi\right)  \right)  , \label{ac.08}%
\end{equation}
and%
\begin{equation}
T_{\mu\nu}^{m}=\left(  \rho_{m}+p_{m}\right)  u_{\mu}u_{\nu}+p_{m}g_{\mu\nu},
\label{ac.09}%
\end{equation}
in which $u_{\mu}=\frac{1}{N}\delta_{t}^{\mu},~u_{\mu}u^{\nu}=-1$, is the
comoving observer.

Moreover, we assume that the scalar field and the matter component share the
symmetries of the FLRW geometry, which implies $\phi= \phi\left(  t \right)  $
and $\rho_{m} = \rho_{m}\left(  t \right)  $. Therefore, for the FLRW
spacetime, the Einstein's field equations are
\begin{equation}
3H^{2} = \frac{\varepsilon}{2N^{2}} \dot{\phi}^{2} + V(\phi) + \rho_{m},
\label{ac.10}%
\end{equation}
\begin{equation}
-\frac{2}{N} \dot{H} - 3H^{2} = \frac{\varepsilon}{2N^{2}} \dot{\phi}^{2} -
V(\phi) + p_{m}, \label{ac.11}%
\end{equation}
where $H = \frac{1}{N} \frac{\dot{a}}{a}$ is the Hubble function, defined by
the expansion rate $\theta= 3H = \nabla_{\mu} u^{\mu}$ for the comoving
observer, and a dot denotes the total derivative with respect to the parameter
$t$, i.e., $\dot{\phi} = \frac{d\phi}{dt}$.

Finally, from the Bianchi identity, $\nabla_{\nu}G^{\mu\nu}=0$, it follows%
\begin{equation}
\nabla_{\nu}\left(  T^{\phi~\mu\nu}+T^{m~\mu\nu}\right)  =0, \label{ac.11a}%
\end{equation}
that is, $\nabla_{\nu}T^{\phi~\mu\nu}=0$, $\nabla_{\nu}T^{m~\mu\nu}=0$, or
equivalently for the FLRW geometry%
\begin{equation}
\frac{\varepsilon}{N}\left(  \frac{\dot{\phi}}{N}\right)  ^{\cdot}%
+\frac{3\varepsilon}{N}H\dot{\phi}+V_{,\phi}=0, \label{ac.12}%
\end{equation}%
\begin{equation}
\frac{1}{N}\dot{\rho}_{m}+3H\left(  \rho_{m}+p_{m}\right)  =0. \label{ac.14}%
\end{equation}

Hence, the equation of state parameter for the scalar field is defined as%
\begin{equation}
w_{\phi}=\frac{\varepsilon\dot{\phi}^{2}-2N^{2}V(\phi)}{\varepsilon\dot{\phi
}^{2}+2N^{2}V(\phi)}, \label{ac.15}%
\end{equation}
while the equation of state parameter for the effective cosmological fluid is
defined as%
\begin{equation}
w_{eff}=\frac{\varepsilon\dot{\phi}^{2}-2N^{2}V(\phi)+2N^{2}p_{m}}%
{\varepsilon\dot{\phi}^{2}+2N^{2}V(\phi)+2N^{2}\rho_{m}}. \label{ac.16}%
\end{equation}

An important characteristic of the cosmological field equations presented
above is that they admit a minisuperspace formulation. Specifically, the
variation of the following Lagrangian function
\begin{equation}
\mathcal{L}\left(  a, \dot{a}, \phi, \dot{\phi} \right)  = -\frac{3}{N} a
\dot{a}^{2} + \frac{\varepsilon}{2N} a^{3} \dot{\phi}^{2} - N a^{3} V\left(
\phi\right)  - N \rho_{m0} a^{-3w_{m}}\,, \label{ac.17}%
\end{equation}
leads to the gravitational model under consideration, where from (\ref{ac.14})
we have substituted $\rho_{m}\left(  a \right)  = \rho_{m0} a^{-3\left(  1 +
w_{m} \right)  }$.

From (\ref{ac.17}), we define the canonical momenta $p_{a}$ and $p_{\phi}$ as
\begin{equation}
\frac{1}{N} \dot{a} = -\frac{1}{6a} p_{a}~, \quad\frac{1}{N} \dot{\phi} =
\frac{1}{a^{3}} p_{\phi},
\end{equation}
and the field equations can then be described by the Hamiltonian function
\begin{equation}
\mathcal{H} \equiv-\frac{1}{12aN} p_{a}^{2} + \frac{\varepsilon}{2a^{3}N}
p_{\phi}^{2} + N a^{3} V\left(  \phi\right)  + N \rho_{m0} a^{-3w_{m}} = 0.
\label{ac.c}%
\end{equation}

We can interpret the field equations as the equations of motion for the point
particles $a$ and $\phi$, evolving within the line element
\begin{equation}
ds^{2} = -6a \, da^{2} + \varepsilon a^{3} d\phi^{2},
\end{equation}
under the influence of an effective potential $U\left(  a, \phi\right)  =
a^{3} V\left(  \phi\right)  + \rho_{m0} a^{-3w_{m}}$. The lapse function is a
non-essential parameter and is introduced solely to enforce the constraint
equation $\mathcal{H} = 0$.

\section{Chameleon Mechanism}

\label{sec3}

Consider now the modified matter Action Integral $\bar{S}_{m}$, with%
\begin{equation}
S_{m}=-\int\sqrt{-g}d^{4}x~f\left(  \phi\left(  x^{\nu}\right)  \right)
L_{m}\left(  \psi\left(  x^{\nu}\right)  ,\nabla_{\mu}\psi\left(  x^{\nu
}\right)  \right)  ,
\end{equation}
where the coupling function $f\left(  \phi\left(  x^{\nu}\right)  \right)  $
describes the energy transfer between the matter source and the scalar field.
From the Bianchi Identity (\ref{ac.11a}) it follows that $\nabla_{\nu}%
T^{\phi~\mu\nu}\neq0,\nabla_{\nu}\bar{T}^{m~\mu\nu}\neq0,$ with $\bar{T}%
_{\mu\nu}^{m}=f\left(  \phi\right)  T_{\mu\nu}^{m}$.

In particular an interaction term is introduced such that%
\begin{equation}
\nabla_{\nu}T^{\phi~\mu\nu}=-Q~,~\nabla_{\nu}T^{m~\mu\nu}=Q, \label{ac.18}%
\end{equation}
such that equation (\ref{ac.11a}) to be satisfied.

For the FLRW background, the cosmological field equations
\begin{equation}
3H^{2}=\frac{\varepsilon}{2N^{2}}\dot{\phi}^{2}+V(\phi)+f\left(  \phi\right)
\rho_{m},
\end{equation}%
\begin{equation}
-\frac{2}{N}\dot{H}-3H^{2}=\frac{\varepsilon}{2N^{2}}\dot{\phi}^{2}%
-V(\phi)+f\left(  \phi\right)  p_{m},
\end{equation}
while equations (\ref{ac.18}) can be written as%
\begin{equation}
\frac{\varepsilon}{N}\left(  \frac{\dot{\phi}}{N}\right)  ^{\cdot}%
+\frac{3\varepsilon}{N}H\dot{\phi}+V_{,\phi}+\left(  1+\alpha\right)
f_{,\phi}\rho_{m}=0,
\end{equation}%
\begin{equation}
\frac{1}{N}\dot{\rho}_{m}+3H\left(  \rho_{m}+p_{m}\right)  -\alpha\dot{\phi
}\left(  \ln f\right)  _{,\phi}\rho_{m}=0,
\end{equation}
where we have considered $Q=\alpha\dot{\phi}\left(  \ln f\right)  _{,\phi}%
\rho_{m}$. Without loss of generality we can assume parameter $\alpha$ to be one.

In the minisuperspace description, the \ point-like Lagrangian (\ref{ac.17})
is modified as
\begin{equation}
\mathcal{\bar{L}}\left(  a,\dot{a},\phi,\dot{\phi}\right)  =-\frac{3}{N}%
a\dot{a}^{2}+\frac{\varepsilon}{2N}a^{3}\dot{\phi}^{2}-Na^{3}V\left(
\phi\right)  -N\rho_{m0}f\left(  \phi\right)  a^{-3w_{m}}. \label{el.001}%
\end{equation}

Consequently, in the Hamiltonian formalism the field equations are described
by the function
\begin{equation}
\mathcal{\bar{H}\equiv}-\frac{1}{12aN}p_{a}^{2}+\frac{\varepsilon}{2a^{3}%
N}p_{\phi}^{2}+Na^{3}V\left(  \phi\right)  +N\rho_{m0}f\left(  \phi\right)
a^{-3w_{m}}=0.
\end{equation}
Thus, the effective potential term in the point description is modified as
$\bar{U}\left(  a,\phi\right)  =a^{3}V\left(  \phi\right)  +\rho
_{m0}a^{-3w_{m}}$.

The definition of the coupling function $f\left(  \phi\right)  $ is essential
for the Chameleon mechanism. From a theoretical point of view, within the
framework of Weyl Integrable theory, function $f\left(  \phi\right)  $ is
considered to be exponential. In the following Section we determine the
functional forms of the potential $V\left(  \phi\right)  $ and the coupling
$f\left(  \phi\right)  $, where the solution space for this model can be
mapped to the solution of the model without interaction, that is $f\left(
\phi\right)  =1$. ~

That is, we answer to the question if there exist a relation $\left(
N,a,\phi\right)  \Longleftrightarrow\left(  \bar{N},\bar{a},\bar{\phi}\right)
$, where the potential functions $U\left(  a,\phi\right)  \Longleftrightarrow
\bar{U}\left(  \bar{a},\bar{\phi}\right)  $.

In order to determine such relation we introduce an external minisuperspace by
using the Eisenhart-Duval lift.

\section{Introducing Interaction via the Eisenhart-Duval lift}

\label{sec4}

The Eisenhart-Duval lift is an approach to the geometrization of Hamiltonian
systems. For dynamical systems of the form $\mathcal{\tilde{H}}=K+U$, where
$K$ is the kinetic energy and $U$ is the potential term, an extended geometry
(lift) is constructed that encodes the potential term. The extended
Hamiltonian, $\mathcal{\tilde{H}}_{\text{lift}} = K_{\text{lift}}$, contains
only a kinetic term; that is, there exists a set of geodesic equations that
describe the original system. More details on the Eisenhart-Duval lift are
presented in Appendix \ref{appen1}.

Using the Eisenhart-Duval lift, it was shown in \cite{comsol} that, for a
class of gravitational models invariant under a certain Lie group of
transformations, there exists a set of canonical transformations that relate
the solutions of these gravitational models. The field equations of such
models can be mapped to an equivalent Hamiltonian system describing the motion
of a particle in a three-dimensional flat geometry.

This property was also found to hold for the scalar field model in vacuum
\cite{comsol2}. In particular, the cosmological field equations with an
exponential potential $V\left(  \phi\right)  =V_{0}e^{-\lambda\phi}$ and
$\rho_{m0}=0$ can be mapped to the equations of motion of a particle in a
three-dimensional flat geometry.

Inspired by the above discussion, we extend the analysis of \cite{comsol2} by
introducing (a) an ideal gas as a matter source and (b) a nonzero interaction
term between the ideal gas and the scalar field. In what follows, we assume
that $\left\vert w_{m}\right\vert < 1$.

\subsection{Scalar field with an ideal gas}

For the Hamiltonian system $\mathcal{H}$, given by (\ref{ac.c}), we introduce
the corresponding extended Hamiltonian function
\begin{equation}
\mathcal{H}_{lift}\mathcal{\equiv}\frac{1}{N^{2}}\left(  -\frac{1}{12a}%
p_{a}^{2}+\frac{\varepsilon}{2a^{3}}p_{\phi}^{2}+\left(  a^{3}V\left(
\phi\right)  +\rho_{m0}a^{-3w_{m}}\right)  p_{z}^{2}\right)  =0, \label{el.01}%
\end{equation}
which describes the geodesic equations for the extended minisuperspace metric
with line element%
\begin{equation}
ds^{2}=-6ada^{2}+\varepsilon a^{3}d\phi^{2}+\frac{1}{a^{3}V\left(
\phi\right)  +\rho_{m0}a^{-3w_{m}}}dz^{2}. \label{el.02}%
\end{equation}

The requirement that the extended Hamiltonian $\mathcal{H}_{\text{lift}}$
{describes the geodesic equations of a conformally flat space implies
that}%

\begin{equation}
V_{A}\left(  \phi\right)  =V_{0}e^{\lambda\left(  \phi-\phi_{0}\right)
}~,~\lambda=\pm\frac{\sqrt{6\varepsilon}}{2}\left(  1+w_{m}\right)
,\label{el.03}%
\end{equation}
It is interesting to note that we recover the exponential potential previously
derived in \cite{comsol2} for the vacuum case. However, now, in the presence
of matter, there exists a relation between the exponential index $\lambda$ and
the equation-of-state parameter $w_{m}$. We next present the transformation
that linearizes the cosmological field equations. {At this point it is
important to remark that for other forms of the Eisenhart lift \cite{ng1} ,
another scalar field potential could arise. However, this investigation is
outside the scopus of this study. }

For potential $V_{A}\left(  \phi\right)  $, and $\lambda=-\frac{\sqrt
{6\varepsilon}}{2}\left(  1+w_{m}\right)  $, we have the transformation%
\begin{equation}
a=L\left(  u\right)  ^{A_{1}}v^{A_{2}},~e^{-\sqrt{\varepsilon}\left(
\phi-\phi_{0}\right)  }=L\left(  u\right)  ^{A_{3}}v^{A_{4}}, \label{el.04}%
\end{equation}
with~%
\begin{align}
A_{1}  &  =\frac{1}{3\left(  3+w_{m}\right)  },~A_{2}=\frac{1}{3\left(
1-w_{m}\right)  },~\label{el.04a}\\
A_{3}  &  =\frac{\sqrt{6}}{3\left(  3+w_{m}\right)  },~A_{4}=\sqrt{\frac{2}%
{3}}\frac{1}{1-w_{m}},~ \label{el.04b}%
\end{align}
and~%
\begin{equation}
V_{0}L\left(  u\right)  +\frac{\left(  3+w_{m}\right)  }{1-w_{m}}\rho
_{m0}L\left(  u\right)  ^{-1+\frac{4}{3+w_{m}}}=-\frac{3}{8}\left(
1-w_{m}\right)  \left(  3+w_{m}\right)  u,
\end{equation}
Thus, the extended minisuperspace reads%
\begin{equation}
ds^{2}=n\left(  u,v\right)  \left(  \dot{u}\dot{v}+\dot{z}^{2}\right)  ,
\label{el.05}%
\end{equation}
where now $n\left(  u,v\right)  =\left(  \frac{v^{\frac{w_{m}}{1-w_{m}}%
}L\left(  u\right)  ^{-2+\frac{1}{3+w_{m}}}}{\rho_{m0}+V_{0}L\left(  u\right)
^{\frac{2\left(  1+w_{m}\right)  }{3+w_{m}}}}\right)  $.

Under the change of variables $\left(  u,v\right)  \rightarrow\frac{1}%
{\sqrt{2}}\left(  X+Y,X-Y\right)  $, the extended Hamiltonian can be written
in diagonal form as
\begin{equation}
\mathcal{H}_{\text{lift}}^{A}\equiv\frac{1}{2N^{2}n\left(  X,Y\right)
}\left(  P_{X}^{2}-P_{Y}^{2}+P_{Z}^{2}\right)  =0, \label{el.06}%
\end{equation}
which describes the geodesic equations of a free particle with zero energy. We
consider $N^{2}n\left(  X,Y\right)  =1$, so that $\mathcal{H}_{\text{lift}%
}^{A}$ is further simplified to $\mathcal{H}_{\text{lift}}^{A}\equiv\frac
{1}{2}\left(  P_{X}^{2}-P_{Y}^{2}+P_{Z}^{2}\right)  =0$. We recall that when
$\rho_{m0}=0$ and $\varepsilon=+1$, the transformation (\ref{el.04}) reduces
to the form presented in \cite{comsol2}, which leads to the linearization of
the exponential potential in the vacuum.

The linearization of the cosmological field equations is achieved due to the
existence of hidden symmetries in the original scalar field model. Indeed, the
original Lagrangian (\ref{el.001}) for the specific exponential potential
admits three symmetries which form the $D\otimes_{s}T_{2}$ Lie algebra
\cite{comsol2}. However, the symmetry vectors of (\ref{el.06}) for a
ten-dimensional Lie algebra. They are the ten conformal symmetries of the
three-dimensional flat space. In the diagonal coordinates where the
Hamiltonian (\ref{el.06}) is epxreesed, they are the six isometries of the
flat space, the one homothetic symmetry and the three proper conformal
symmetries. In the coordinates $X,Y,Z$ the symmetry vectors are
\[
\partial_{X},~\partial_{Y},~\partial_{Z},
\]%
\[
Y\partial_{X}+X\partial_{Y}~,~Z\partial_{X}-X\partial_{Z}~,~Z\partial
_{Y}+Y\partial_{Z},
\]%
\[
X\partial_{X}+Y\partial_{Y}+Z\partial_{Z},
\]%
\[
\frac{X^{2}+Y^{2}-Z^{2}}{2}\partial_{X}+XY\partial_{Y}+XZ\partial_{Z},
\]%
\[
XY\partial_{X}+\frac{Y^{2}+X^{2}+Z^{2}}{2}\partial_{Y}+YZ\partial_{Z},
\]%
\[
XZ\partial_{X}+YZ\partial_{Y}+\frac{Z^{2}+Y^{2}-X^{2}}{2}\partial_{Z}.
\]
{These vector fields are symmetries of the extended minisuperspace, and
they should not be confused with the structure of the background geometry. }

\subsection{Chameleon Mechanism}

For the cosmological field equations with the Chameleon Mechanism, described
by the Hamiltonian function $\mathcal{\bar{H}}$, we introduce the
corresponding extended Hamiltonian.
\begin{equation}
\mathcal{\bar{H}}_{lift}\mathcal{\equiv}\frac{1}{N^{2}}\left(  -\frac{1}%
{12a}p_{a}^{2}+\frac{\varepsilon}{2a^{3}}p_{\phi}^{2}+\left(  a^{3}V\left(
\phi\right)  +\rho_{m0}a^{-3w_{m}}\right)  p_{z}^{2}\right)  =0. \label{el.07}%
\end{equation}

Similarly to before, the cosmological field equations are described by the
geodesic equations for the extended minisuperspace metric with line element%
\begin{equation}
d\bar{s}^{2}=-6ada^{2}+\varepsilon a^{3}d\phi^{2}+\frac{1}{a^{3}V\left(
\phi\right)  +\rho_{m0}f\left(  \phi\right)  a^{-3w_{m}}}dz^{2}. \label{el.08}%
\end{equation}

Hence, we calculate the Cotton-York tensor for the latter three-dimensional
metric, we found that it is conformally flat if and only if
\begin{equation}
V_{A}\left(  \phi\right)  =V_{0}e^{\lambda\left(  \phi-\phi_{0}\right)
},~f_{A}\left(  \phi\right)  =V_{0}e^{-2\lambda\left(  \phi-\phi_{0}\right)
},~~\lambda=\pm\frac{\sqrt{6\varepsilon}}{2}\left(  1+w_{m}\right)  .
\label{el.09}%
\end{equation}

Therefore under the change of variables
\begin{equation}
a=u^{A_{1}}M\left(  v\right)  ^{A_{2}},~e^{-\sqrt{\varepsilon}\left(
\phi-\phi_{0}\right)  }=u^{A_{3}}M\left(  v\right)  ^{A_{4}},
\end{equation}
where $A_{1}$,~$A_{2}$,~$A_{3}$ and $A_{4}$ are given by expressions
(\ref{el.04a}), (\ref{el.04b}), and~$\left(  u,v\right)  \rightarrow\frac
{1}{\sqrt{2}}\left(  X+Y,X-Y\right)  $, the extended Hamiltonian reads
\begin{equation}
\mathcal{\bar{H}}_{lift}^{A}\mathcal{\equiv}\frac{1}{2N^{2}\bar{n}\left(
X,Y\right)  }\left(  P_{X}^{2}-P_{Y}^{2}+P_{Z}^{2}\right)  ,
\end{equation}
where $\bar{n}\left(  u,v\right)  =\frac{u^{-1+\frac{1}{3+w_{m}}}M\left(
v\right)  ^{\frac{w_{m}}{1-w_{m}}}}{V_{0}+\rho_{m0}M\left(  v\right)
^{2-\frac{4}{1-w_{m}}}}$, and function $M\left(  v\right)  $ is given as
follows%
\begin{equation}
V_{0}M\left(  v\right)  -\frac{1-w_{m}}{1+3w_{m}}\rho_{m0}M\left(  v\right)
^{3-\frac{4}{1-w_{m}}}=-\frac{3}{8}\left(  1-w_{m}\right)  \left(
3+w_{m}\right)  .
\end{equation}

Therefore by selecting $N^{2}\bar{n}\left(  X,Y\right)  =1$, the extended
Hamiltonian is simplified as $\mathcal{\bar{H}}_{lift}^{A}\mathcal{\equiv
}\frac{1}{2}\left(  P_{X}^{2}-P_{Y}^{2}+P_{Z}^{2}\right)  $.

\section{Conclusions}

\label{sec5}

In this study, we considered a scalar field cosmology with a perfect fluid
characterized by a constant equation-of-state parameter, with or without an
interaction term between the matter components of the cosmic fluid. The
cosmological field equations admit a minisuperspace formulation and correspond
to constrained Hamiltonian systems described by second-order differential equations.

We employed the Eisenhart-Duval lift and presented an equivalent Hamiltonian
system that describes the geodesic equations for an extended minisuperspace
metric. It was found that, for a power-law potential $V\left(  \phi\right)  $
and either a constant interaction or an exponential coupling function
$f\left(  \phi\right)  $, there exist canonical transformations within the
extended minisuperspace that allow the field equations to be expressed as null
geodesic equations for a free particle in flat geometry. This means that there
exists a canonical transformation through which one can construct the analytic
solution of the model with a nonconstant coupling function $f\left(
\phi\right)  $, starting from the model with a constant coupling function.

Furthermore, it was shown in \cite{comsol2} that the scalar field with an
exponential potential in vacuum exhibits the same property. That is, there
exists a transformation such that the field equations can be written in terms
of a three-dimensional free particle system. Recently, in \cite{anlambda},
this equivalence with the three-dimensional system was also found for the
field equations of the Cold Dark Matter (CDM) and $\Lambda$CDM models. The
equivalence follows from the application of the Lorentzian lift to the field
equations of the $\Lambda$CDM model. In the extended minisuperspace, there
exist canonical transformations that relate the solution spaces of the two
cosmological models and allow us to construct the solution of the one model
from the other solution.

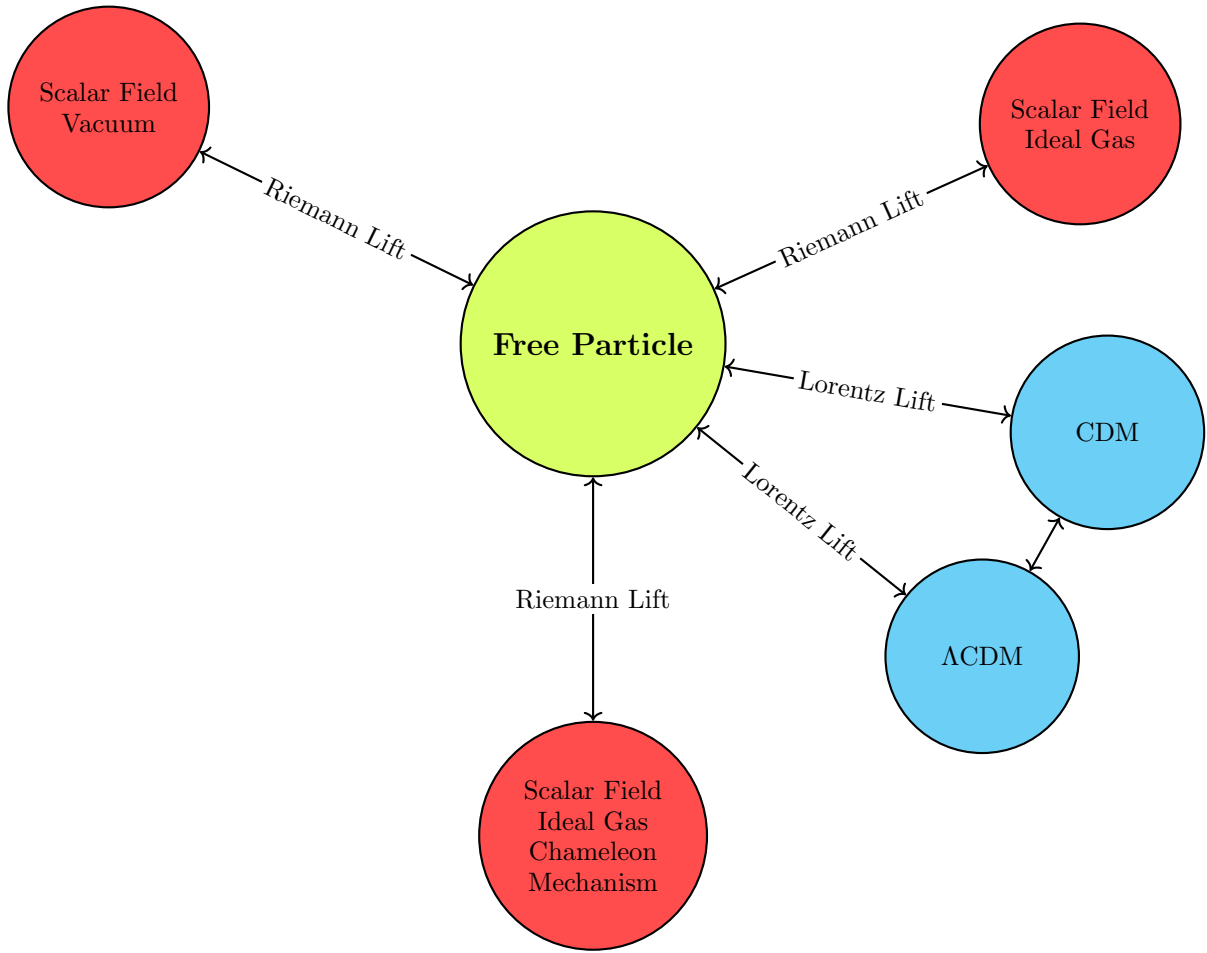
\begin{figure}[ptbh]
\centering
\begin{tikzpicture}[
every node/.style={align=center},
arrow/.style={<->, thick},
comment/.style={font=\normalsize, midway, sloped, fill=white, inner sep=2pt},
center node/.style={circle, draw=black, thick, fill=lime!60, minimum size=3.5cm, text width=3cm, font=\bfseries\large, align=center},
outer node/.style={circle, draw=black, thick, minimum size=2.5cm, text width=2.3cm, font=\normalsize, align=center},
scalar/.style={outer node, fill=red!70},
cosmology/.style={outer node, fill=cyan!50}
]
\node[center node] (free) at (0,-2) {Free Particle};
\node[scalar] (vacuum) at (170:6.5cm) {Scalar Field\\Vacuum};
\node[scalar] (ideal) at (8:6.5cm) {Scalar Field\\Ideal Gas};
\node[scalar] (chameleon) at (-90:8.5cm) {Scalar Field\\Ideal Gas\\Chameleon Mechanism};
\node[cosmology] (cdm) at (-25:7.5cm) {CDM};
\node[cosmology] (lcdm) at (-50:8cm) {$\Lambda$CDM};
\draw[arrow] (free) to[bend left=0] node[comment] {Riemann Lift} (vacuum);
\draw[arrow] (free) to[bend right=0] node[comment] {Riemann Lift} (ideal);
\draw[arrow] (free) -- node[comment, rotate=90] {Riemann Lift} (chameleon);
\draw[arrow] (free) -- node[comment] {Lorentz Lift} (cdm);
\draw[arrow] (free) -- node[comment] {Lorentz Lift} (lcdm);
\draw[arrow] (cdm) -- (lcdm);
\end{tikzpicture}
\caption{Canonical structure for the solution space between the different
cosmological models}%
\label{fig0}%
\end{figure}

The field equations of the three-dimensional free particle appear to be a
common characteristic of the gravitational models discussed above. Thus, the
solution spaces of these models are equivalent, and starting from the free
particle, one can construct all the corresponding cosmological solutions. In
Fig.~\ref{fig0}, we present this connection in diagrammatic form. This
dynamical equivalence holds only at the background level and is lost in the
case of cosmological perturbation theory. This is because the minisuperspace
description is no longer valid when perturbations are introduced.

Last but not least, we emphasize that the equivalence discussed here applies
to the solution spaces of Einstein's field equations and
should not be confused with the physical equivalence of different cosmological
models. Indeed, such models may possess different degrees of freedom and fit
observational data in distinct ways. Specifically, in~\cite{chd}, the
chameleon model with an exponential potential was tested using late-time
cosmological data. Compared to the $\Lambda$CDM model, it yielded different
values for the physical parameters as well as distinct statistical measures.
However, the issue of cosmological tensions was not addressed in~\cite{cosm1}.
Whether the Eisenhart-Duval lift can be applied to tackle these
tensions remains a subject for further investigation, which lies beyond the
scope of the present work and will be addressed in future research.

{It is important to mention that the Eisenhart-Duval lift is not the
unique approach to derive this kind of hidden symmetries which lead to this
family of canonical transformations. An alternative approach has been
established in a series of studies \cite{jb1,jb2,jb3,jb4}, where the
derivation of the hidden symmetries was performed from the application of
Noether's condition. In contrary to this study, where the existence of the
hidden symmetries and of the linearized properties where derived using only
geometric tools,  as the Cotton-York tensor, which provide information about
the geometric structure of the extended minisuperspace. } Last but not least,
it is important to mention that another class of canonical transformations
which relate scalar field inflationary solutions is presented in \cite{bar1}.

\begin{acknowledgments}
The author thanks the support of Vicerrector\'{\i}a de Investigaci\'{o}n y
Desarrollo Tecnol\'{o}gico (Vridt) at Universidad Cat\'{o}lica del Norte
through N\'{u}cleo de Investigaci\'{o}n Geometr\'{\i}a Diferencial y
Aplicaciones, Resoluci\'{o}n Vridt No - 096/2022. AP was partially supported
by Proyecto Fondecyt Regular 2024, Folio 1240514, Etapa 2025.
\end{acknowledgments}

%

\appendix

\section{Eisenhart-Duval Lift}

\label{appen1}

The Eisenhart-Duval lift provides a systematic methodology for the
geometrization of dynamical systems \cite{ll01,ll02}. This geometrization is
achieved by reformulating a given dynamical system as an equivalent set of
geodesic equations in a higher-dimensional space. A necessary requirement is
the existence of a sufficient number of conservation laws for the geodesic
equations in order to reduce them and recover the original dynamical system. A
detailed and pedagogical discussion on the application of the Eisenhart-Duval
lift is presented in \cite{ll03}. In the following lines, we present the basic definitions.

Assume a dynamical system which follows from the variation of the following
Lagrangian function%
\begin{equation}
\tilde{L}\left(  N,q^{k},\dot{q}^{k}\right)  =\frac{1}{2}g_{ij}\dot{q}^{i}%
\dot{q}^{j}-V(q^{k}). \label{sl.09}%
\end{equation}
This Lagrangian describes the equations of motion of a particle in a curved
manifold with $n-$dimensional space with metric tensor $g_{ij}$ under the
action of the autonomous potential $V\left(  q^{k}\right)  $

We derive the momentum $p_{i}=g_{ij}\dot{q}^{j}$ and we write the Hamiltonian
function
\begin{equation}
H\left(  q^{k},p^{k}\right)  \equiv\frac{1}{2}g^{ij}p_{i}p_{j}+V(q^{k})=h,
\label{sl.10}%
\end{equation}
where $h$ is the value of the conservation law. The corresponding equations of
motion are
\begin{align}
\dot{q}^{i}  &  =g^{ij}\dot{p}_{j},\label{sl.11}\\
\dot{p}_{k}  &  =-\frac{1}{2}g_{~,k}^{ij}p_{i}p_{j}-V_{,k}. \label{sl.12}%
\end{align}

Consider now the extended Hamiltonian%
\begin{equation}
H_{n+1}=\frac{1}{2}g^{ij}p_{i}p_{j}+\frac{1}{2}V(q^{k})p_{z}^{2}=h_{n+1},
\label{sl.14}%
\end{equation}
where now describes the geodesic equations for the extended metric%
\begin{equation}
ds_{n+1}^{2}=g_{ij}\left(  q^{k}\right)  dq^{i}dq^{j}+\frac{1}{V\left(
q^{k}\right)  }dz^{2}. \label{sl.19}%
\end{equation}
The geodesic equations are
\begin{align}
\dot{q}^{i}  &  =g^{ij}\dot{p}_{j},\label{sl.15}\\
\dot{z}  &  =\frac{1}{V\left(  q^{k}\right)  }p_{z},\label{sl.16}\\
\dot{p}_{k}  &  =-\frac{1}{2}g_{~,k}^{ij}p_{i}p_{j}-\frac{1}{2}V_{,k}p_{z}%
^{2},\label{sl.17}\\
\dot{p}_{z}  &  =0. \label{sl.18}%
\end{align}
We observe that $p_{z}$ is conserved, which follows directly from the isometry
$\partial_{z}$ of the extended metric tensor. Hence, by substituting this
conservation law into the remaining equations of motion, we recover the
original dynamical system. This approach is known as the Riemann lift.

However, this is not the only way to perform the lift. The Lorentz lift is a
widely used alternative, in which the dynamical system is expressed in terms
of geodesic equations for a metric spacetime written in Brinkmann coordinates,
similar to those used in pp-wave spacetimes.

The extended Hamiltonian
\begin{equation}
H_{n+2}\equiv\frac{1}{2}g^{ij}p_{i}p_{j}+V(q^{k})p_{u}^{2}+p_{u}p_{v}=h_{n+2},
\end{equation}
with equations of motion%
\begin{align}
\dot{q}^{i}  &  =g^{ij}\dot{p}_{j},\\
\dot{u}  &  =2V\left(  q^{k}\right)  p_{u}+p_{v},\\
\dot{v}  &  =p_{u},\\
\dot{p}_{k}  &  =-\frac{1}{2}g_{~,k}^{ij}p_{i}p_{j}-V_{,k}p_{u}^{2},\\
\dot{p}_{u}  &  =0,\\
\dot{p}_{v}  &  =0.
\end{align}
and $p_{u}$,~$p_{v}$ to be conservations laws related to the isometries
$\partial_{u}$ and $\partial_{v}$ for the extended metric tensor%
\begin{equation}
ds_{\left(  n+2\right)  }^{2}=g_{ij}\left(  q^{k}\right)  dq^{i}%
dq^{j}+2dudv-2V\left(  q^{k}\right)  du^{2}.
\end{equation}

In this study, we focused on the Riemann and Lorentz lifts; however, for
alternative lifting approaches, we refer the reader to \cite{ng1}.

\end{document}